\documentstyle[11pt,newpasp,epsfig,twoside]{article}
\def\lax    {\ifmmode{_<\atop^{\sim}}\else{${_<\atop^{\sim}}$}\fi}
\def\gax    {\ifmmode{_>\atop^{\sim}}\else{${_>\atop^{\sim}}$}\fi}

\newfont{\ssq}{cmssq8}

\def\edcomment#1{\iffalse\marginpar{\raggedright\sl#1\/}\else\relax\fi}
\reversemarginpar

\begin{document}
\title{Physical Modelling of the Zodiacal Dust Cloud}

\author{\bf Leonid M. Ozernoy}

\affil{5C3, School of Computational Sciences and Department of Physics
\& Astronomy, George Mason U., Fairfax, VA 22030-4444, USA}

\begin{abstract}

This review is based on an extensive work done in collaboration with
N.~Gorkavyi, J.~Mather, and T.~Taidakova, which aimed at the physical
modelling of the interplanetary dust (IPD) cloud in the Solar system,
{\it i.e.}, establishing a link between the observable characteristics
of the zodiacal cloud and the dynamical and physical properties of
the parent minor bodies. Our computational approach permits with modest
computer resources to integrate the trajectories of hundreds of
particles and to effectively store up to $10^{10-11}$ positions, which
provides a high fidelity 3D distribution of the dust.
Our numerical codes account for the major dynamical effects that govern
the motion of IPD particles: the Poynting-Robertson (P-R) drag and solar
wind drag; the solar radiation pressure; particle evaporation;
gravitational scattering by the planets; and the influence of
mean-motion resonances. The incorporation of secular resonances and
collisions of dust particles (both mutual and with interstellar dust) is
underway.  We have demonstrated the efficacy of our codes by performing
the following analyses: (i) simulation of the distribution of Centaurs
(comets scattered in their journey from the Kuiper belt inward the Solar
system) and revealing the effects of the outer planets in producing
`cometary belts'; (ii) detailed inspection of a rich resonant structure found
in these belts, which predicts the existence of gaps similar to the Kirkwood
gaps in the main asteroid belt;  (iii) a preliminary 3-D physical model of
the IPD cloud, which includes three dust components -- asteroidal, cometary,
and kuiperoidal -- and is consistent with the available data of Pioneers
and Voyagers dust detectors;  (iv) modeling of the IPD cloud, which
provides a zodiacal light distribution in accord, to the order of 1\%, with
a subset of the COBE/DIRBE observations; and (v) showing that the resonant 
structure in dusty circumstellar disks of Vega and Epsilon Eridani is a 
signature of embedded extrasolar planets. Further improvements of our 
modelling and their importance for astronomy and cosmology are outlined.
\end{abstract}

\section*{1. Introduction}

The necessity to account accurately for zodiacal emission as part of the 
measurements of the cosmic infrared background  has revived and strengthened
a long-standing interest in the theoretical aspects of the dynamics, 
structure, and evolution of the interpanetary dust (IPD) cloud ({\it e.g.},
Hauser et al. 1998).
Until recently, the main stumbling block to implementing the comprehensive
study of IPD has been the absence of a physical model for the interplanetary
dust cloud. Such a model would establish a link between the observable
characteristics of the zodiacal cloud and the dynamical and physical
properties of the parent minor bodies of the Solar system. Recently, we
constructed a preliminary physical model of the IPD cloud based on
combining new methods elaborated by the present authors (Gorkavyi, Ozernoy
\& Mather [GOM hereinafter] 1996;  GOM 1997; Gorkavyi et al. [GOMT 
hereinafter] 1997a-c; GOMT 1998a-c, 2000a-c) {\it and} some
features of previously used numerical and analytical methods
(Haug 1958; Kessler 1981; Leinert et al. 1983; 
Dermott et al. 1996; Valsecchi \& Manara 1997; Liou \& Zook 1999).
Specifically, our approach, which is based on the use of the kinetic
equation for the density of dust in the space of orbital coordinates as well
as in the ordinary space, combines analytical (kinetic or hydrodynamical)
methods in conjunction with celestial mechanics orbit calculations and
numerical computations. More recently, we have substantially strengthened
the computational component of our approach by developing a new, more
powerful technique described in the next Section, which enables us to get
rid of a number of approximations used before and therefore considerably
improve our physical modelling.
      
\section*{2. A New Computational Approach Employed}

The development of the physical model of the IPD
cloud is complicated by:
\begin{itemize}     
\item {} 
 uncertainties in the distribution of comets or other bodies as the
     major sources of dust in planetary systems;

\item {}  difficulties in specifying accurately the level and times (positions)
     of dust contribution by the objects; and

\item {}  the wide variety of relevant physical processes, such as (i) radiation
     pressure and dissipative effects (Poynting-Robertson drag and stellar
     wind drag), (ii) resonant interaction with planets, (iii) gravitational
     scattering by planets, (iv) evaporation and sputtering of dust
     particles, (v) mutual collisions in the cometary and dust populations,
     and (vi) orbital variations in the dust production rate by minor bodies.
\end{itemize}

These complexities exclude attaining an analytical solution, so reliance
must be on the numerical modelling. In our approach, two features are worth
mentioning: (1) the techniques employed permit construction of
high-quality 3D models of IPD clouds with the number of particles
(strictly speaking, particle positions) as high as $10^{(10-11)}$; and (2)
incorporation of an original stable numerical integrator suitable for both
dissipationless and, with some modification, dissipational dynamics of minor
bodies or particles.
     
\subsection*{2.1. Our computational approach}

Up until now, numerical models suffered from the limited number of particles
that could be used in the computations. For instance, so far the best results 
in modeling the dynamics of dust from the Kuiper belt were obtained by Liou
\& Zook (1999) with $50-100$ particles of three different sizes using 
$\sim 10^4$ positions for each particle giving $\sim 10^6$ particle positions. 
Those authors used their results for
2-D projection of the zodiacal cloud with resolution 1 AU and typical
statistics 200-400 particles/AU$^2$ (or 20-40 particles/AU$^3$ for a 3-D
model). Such modeling cannot easily match the large-scale structure of the
IPD cloud, the respected  maps of the zodiacal emission, etc.

In our approach described in Ozernoy, Gorkavyi, \& Taidakova [OGT
hereinafter] (2000a,b) and Ozernoy, Gorkavyi, Mather \& Taidakova [OGMT
hereinafter] (2000), the particle-number limitation is substantially
relaxed, which is
decisive to provide reliable numerical simulations. In brief, our approach
is as follows (for simplicity of understanding, we consider a stationary dust
particle distribution in the frame co-rotating with the planet). The locus of
the given  particle's
positions (taken, say, as $6\cdot 10^3$ positions every revolution about the
star) are recorded and considered as the positions of {\it many other
particles} produced by the same source of dust but {\it at a different
time}. After this particle `dies' (as a result of infall or ejection from
the system by a planet-perturber), its recorded positions sampled over its
lifetime form a stationary distribution as if it were produced by {\bf many}
particles. Typically, each run includes  $10^4-10^5$ revolutions, i.e.
 $\sim 10^8$ positions of a dust particle, which is equivalent, for a
stationary distribution, to $10^8$ particles. If we allow for 100 sources of
dust (in fact, we can include, if necessary,  a larger number of sources),
after 100 runs we deal with $\sim 10^{10}$ particle positions as if they
were real particles. In the present project, we will not only keep
information about the dynamical path of each particle (as we did in OGT
2000a,b), but in addition, we will immediately sort the information about
the computed coordinates of each particle into $10^6-10^7$ spatial cells
(each cell containing $10^3-10^4$ particles), thereby forming a 3D grid that
models the dust cloud around the Sun (or a star) (OGMT 2000).
An appreciable increase in statistics, compared to Liou \& Zook (1999),
brings a factor of $10^4$ improvement in the detail of a model and enables
us to model the IPD cloud at a qualitatively new, 3-D level. Moreover, our
approach makes it possible to study, besides stationary processes, certain
non-stationary processes as well, e.g. evolution toward steady-state
distributions, dust production from non-steady sources, decrease in particle
size (due to evaporation and sputtering) and number (due to collisions), etc.
 
\subsection*{2.2. Numerical Integrators for Dissipationless and
Dissipative Systems}

Our computational method as well as the use of an implicit second-order
integrator (Taidakova 1997) appropriately adapted to achieve our goals
(Taidakova \& Gorkavyi 1999) are described in more detail in OGT (2000b);
as shown there, the integrator for a
dissipationless system provides the necessary accuracy of computations on
the time scale of $0.5\cdot 10^9$ years. A big advantage of this integrator
is its stability: an error in the energy (the Tisserand parameter) does not
grow as the number of time steps increases if the value of the step remains
the same. The latter situation is exemplified by a {\it resonant particle}
-- it does not approach too close to the planet so that the same time step
can be taken. Meanwhile {\it non-resonant particles}, in due course of their
gravitational scatterings, approach one or another planet from time to time,
and therefore one has to change the time step near the planet. Obviously,
whenever the time step diminishes near the planet, an error in the Tisserand
parameter slowly grows together with the increased number of the smaller
time steps. Nevertheless, in our simulations a fractional error in the
Tisserand parameter typically does not exceed 0.001 during $3\cdot 10^6$
Neptune's revolution (OGT 2000b), which amounts to 0.5 Gyrs. To increase
the accuracy of the computations, we use a second iteration (OGT 2000a).
While the 1st iteration yields the gravitational field between points $A$
and $B$ using an approximative formula based on the  particle parameters at
point $A$ (because those at point $B$ are still unknown), the 2nd iteration
enables us to compute the gravitational field between $A$ and $B$ using a
middle position between them because the position of $B$ is already given by
the 1st iteration.

As for dissipative systems (e.g., with a P-R drag), a modified implicit
second-order integrator has been elaborated (Taidakova and Gorkavyi 1999).
It provides a necessary accuracy of integration, which remains stable if the
time step of computations is changed in jumps, and not continuously, as the
particle approaches the Sun. By applying this approach, one can compute the
dynamical evolution of dust particles accounting for virtually {\it all}
physical processes listed above. 
\section*{3. Components of the Interplanetary Dust Cloud}

\subsection*{3.1. Asteroidal Component of Dust}

Under the bombardment of other asteroids and large grains, each asteroid
serves as a source of dust. The asteroidal dust gradually  approaches the
Sun due to the P-R drag or escapes from the Solar system due to gravitational
scattering by Jupiter and radiation pressure (for small particles) and
solar wind (for small to moderate sized particles, depending on their
charge, etc.).

The asteroidal component of the IPD has the following features: (i) its
distribution is flat; (ii) the dust density run is expected to be $R^{-1}$
at $R\sim 1$~AU and have a cut-off at $R>2$ AU (GOMT 1997a), $R$ being
heliocentric distance; (iii) the
asteroidal dust is mainly responsible for the content of the Earth resonant
ring (Jackson \& Zook 1989, Dermott et al. 1994).
Using the `dust bands' data, Dermott et al. (1996) estimate
the fraction of asteroidal particles in the IPD cloud to be about 1/3.

\subsection*{3.2. Cometary Component of Dust}

The cometary component of dust originates from sublimation of comets and has
the following features: (i) its distribution is relatively thick; (ii) the
dust density run is expected to be $R^{-2.4}$ at $R\sim 1$~AU (GOMT 1997a);
(iii) most of the cometary dust escapes from our planetary system due to
perturbations by Jupiter and the solar radiation pressure. Except a few
attempts (Liou et al. 1995, GOMT 1997a), so far there are
no reliable estimations of the cometary dust fraction in the IPD cloud.

\subsection*{3.3. Dust From the Kuiper Belt Objects}

The sources of the IPD cloud cannot be entirely reduced simply to
comets (a part of which is also responsible for the observed dust tails)
and to asteroids (a part of which assembled in asteroid families is also
responsible for the observed `dust bands' in the IPD emission)
-- a number of facts force us to suspect that additional sources of
interplanetary dust must exist:

     1. Chemical analyses indicate that a part of IPD spent a much longer
time in space than the typical asteroidal and cometary particles (Flynn
1994, 1996).

     2. Pioneer 10 and 11 data indicate that the dust particles of mass
$10^{-(8-9)}$~g have approximately constant flux seen up to 18 AU (Humes
1980, Divine 1993). Similarly,
 Voyager 1 and 2 data indicate that the dust particles of mass
$10^{-(11-12)}$~g are seen from 6 to 40 AU with approximately constant flux
$(0.5-1)\cdot 10^{-3}$ particles/m$^2$/s (Gurnett et al. 1997).
Both these results cannot be explained by the cometary and asteroidal
sources giving an entirely different distribution of dust
(GOMT 1997a).

3. The total number of KBOs inferred by from available observations is
$8\cdot 10^8$ (Jewitt 1999), which  exceeds the number of the known
Jupiter family comets by a factor of $10^6$.
This indicates that the overall dust production rate from KBOs may be 
not negligible
compared to that of comets and hence  a third important component of
the IPD cloud might be the `kuiperoidal' dust, as it has recently been
suspected (Backman et al. 1995).   

In our opinion, the Kuiper belt influences the formation of the IPD cloud in
two ways: (i)  as a source of small-size particles  slowly drifting toward
the Sun under a combined action of the P-R drag  and perturbations from the
planets; and (ii) as a source of millions of comets between Jupiter and
Neptune (Levison \& Duncan 1997; OGT 2000a,b), which, in turn, serve as
additional sources of dust. The dust can be produced
due to evaporation of the volatile material from the KBO surface as a result
of a variety of processes, such as the Solar wind and the heating by the Sun, 
micrometeor bombardment, mutual collisions of kuiperoids (e.g., Stern 2000),
etc.  Although these processes are very complicated, further work could
enable us to put important constraints on the contribution of kuiperoidal
dust in the overall dust balance.
 
So far there are  no
reliable estimations of the kuiperoidal dust fraction in the inner Solar
system. Our working hypothesis quantitatively analyzed below
is that {\it KBOs and Centaurs} (invisible comets mainly
beyond Jupiter) {\it could produce an important contribution to the dust
content of the IPD cloud}.

\section*{4. Simulating the Distributions of Dust Sources and
 Interplanetary Dust}

\subsection*{4.1. The Simulated Distribution of Dust Sources in the 
Outer Solar System}

The outer Solar system beyond the four giant planets includes the Kuiper
belt and the Oort cloud, which contain raw material left since the formation
of the system. The KBOs are thought to be responsible for progressive
replenishment of the observable cometary populations, and gravitational
scattering of these objects by the four giant planets can provide their
transport from the trans-
\begin{figure} [!ht]
\centerline{\epsfig{file=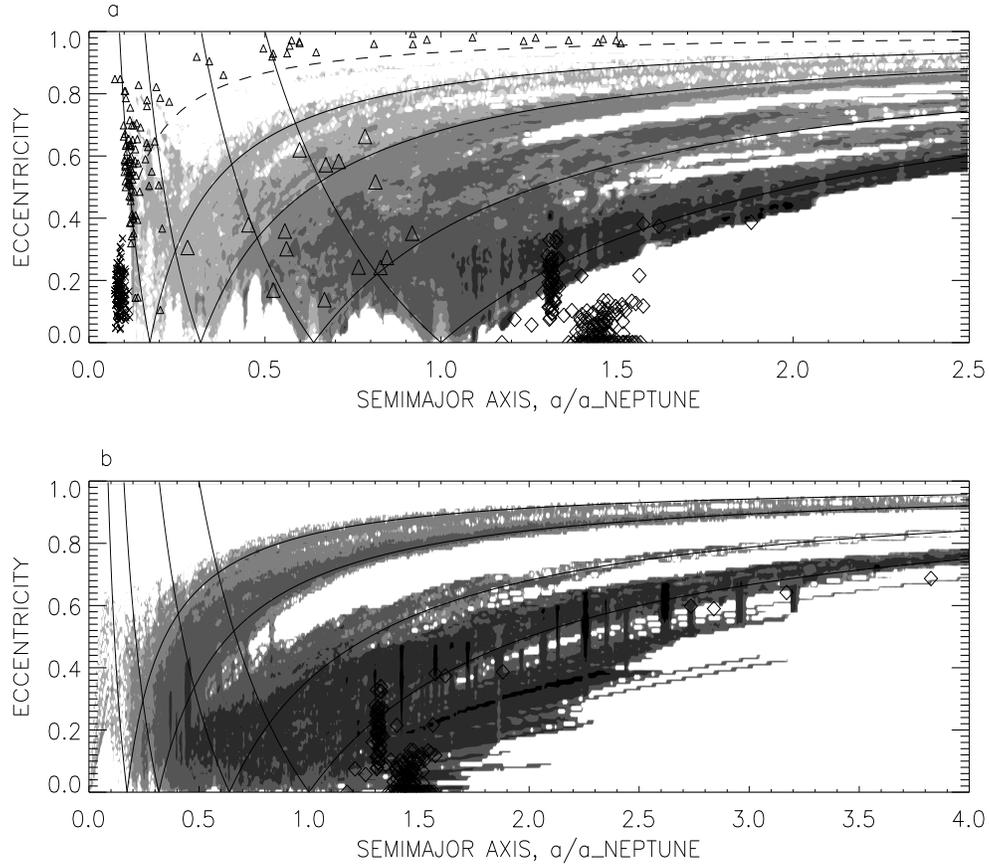,width=5.5in,height=4.7in}}
\caption{\ssq
The simulated cometary and dust populations
in coordinates `eccentricity - semimajor axis'. To represent the number
of comets/dust particles in each cell, a logarithmic scale with 6 grey
levels is used, i.e. each shade differs 10-fold from the neighboring one.
The boundaries of the so called crossing zones (where the orbits of comets 
and planets are crossed) are
shown by heavy lines. Diamonds stand for the (205) Kuiper belt objects.\\
{\bf a} -- 2D density of the simulated cometary population in the outer Solar
system (eight-planet approximation)  (OGT 2000a). Four cometary belts of
the giant planets with resonant structures can be seen. The region occupied
by visible comets (perihelion distances $<2$ AU) is located above the
dashed line. The known asteroids of the main belt (100 objects) are shown
by crosses, small triangles stand for short-periodic comets (112
objects), and large triangles stand for Centaurs (15 objects).\\
{\bf b} -- 2D density of the simulated population of the kuiperoidal
dust (eight-planet approximation, 200 sources) (GOMT 2000a).
Four cirsumsolar dust belts, with
the scattered components along the boundares of each
planet's crossing zone, containing rich resonant structures
can be seen  near the orbits of the giant planets.
}
\label{fig1}
\end{figure}
\noindent
Neptunian region all the way inward, down to
Jupiter (Levison \& Duncan 1997; Malhotra, Duncan \&  Levison 1999; OGT
2000a,b). An approach started in OGT 2000a,b makes emphasis on the structure
of cometary populations between Neptune and Jupiter, both in phase space, 
{\it i.e.}, in the space of orbital coordinates \{$a,e,i$\}, and in real 
space. Using numerical simulations, we examined the structure of a cometary
population near a massive planet, such as a giant planet of the Solar
system, starting with one-planet approximation (the Sun plus one planet). By
studying the distributions of comets in semimajor axis, eccentricity,
pericenter, and apocenter distances, we revealed several interesting
features in these distributions. The most remarkable ones include: (i)
each giant planet dynamically controls a cometary population that we call
the {\it `cometary belt'}); and, (ii) avoidance of resonant orbits by
comets. We then enhanced the calculations by determining how a cometary
belt is modified when the influence of all eight planets is taken into
consideration.  To this end, we simulated a stationary distribution of comets,
which results from the gravitational scattering of the Kuiper belt objects
by mainly four giant planets and takes into account the effects of mean
motion resonances. The objects start from the Kuiper belt and are typically
traced until the bulk of comets ($\sim 90$\%) are ejected from the Solar
system (this happens on a time scale of $\lax 0.5$ Gyrs). Accounting for
the influence of four giant planets makes the simulated cometary belts
overlap (Fig.~1a), but nevertheless keeps almost all their basic features
found in the one-planet approximation. In particular, the simulated belts
maintain the gaps in the $(a,e)$- and $(a,i)$-space similar to the Kirkwood
gaps in the main asteroid belt.

The simulated spatial accumulations of comets near the orbits of all four
giant planets -- the cometary belts -- have a dynamical nature, because the
comets belonging to the given planet's belt are either in a resonance with
the host or are gravitationally scattered predominantly by this planet. We
conclude that the {\it large-scale structure of the Solar system includes the
four cometary belts} expected to contain at present 20-30 million scattered
comets.  Only a tiny fraction of them is currently visible as Jupiter-,
Saturn-, etc. family comets.

\subsection*{4.2. Simulations of Dust Distribution from Kuiper
Belt Objects}

Knowledge of the simulated distribution of sources of dust, along with the
known sources, has enabled us to compute the structure of the asteroidal,
cometary, and kuiperoidal components of the IPD cloud. 
Here, we describe the structure of the latter. 

In accordance with the main dynamical factors, we would expect to get three
major components of the kuiperoidal dust:  i)~`freely' drifting particles,
(ii)~gravitationally scattered particles, and (iii) particles captured into
resonances.

In the phase space, we find the dust distribution highly non-uniform, with
most of the dust concentrating into the four belts associated with the
orbits of the four giant planets, with the Neptune dust belt being the most
dense and extended (Fig.~1b). As distinct from the  simulated cometary belts 
described in Sec.~4.1, for which the dominating gravitational scattering 
results in avoidance of resonant orbits by comets,
the dust belts, due to an additional factor -- the P-R drag -- reveal
a rich and complex resonant structure of captured  particles.

Using our approach, we have reconstructed the spatial
structure of the IPD cloud in the Solar system between 0.5 and 100 AU. Our
simulations offer a 3-D physical model of the kuiperoidal dust cloud based
on $(2-8)\times 10^6$ cells 
\begin{figure} [!htb] 
\centerline{\epsfig{file=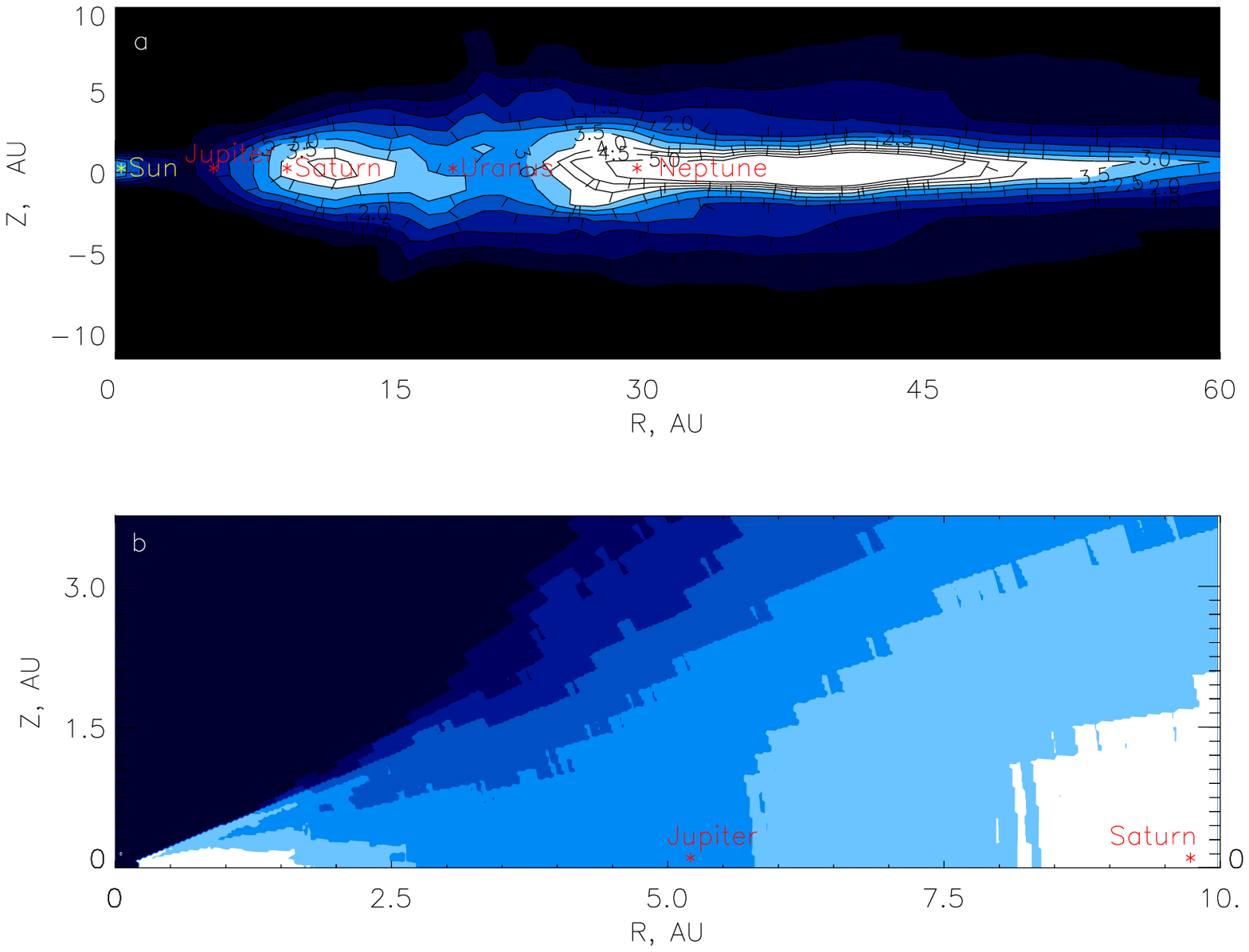,width=5.5in,height=4.1in}}
\caption{\ssq
{\bf a} -- An edge-on view on the simulated distribution of kuiperoidal
dust in the Solar system produced from 133 KBOs. This 3D model
is constructed on a restangular $200\times 200\times 200$ grid with
$8\cdot 10^6$ cells containing $0.5\cdot 10^{11}$ particle positions.\\
{\bf b} -- The simulated distribution of kuiperoidal dust in the inner part
of the Solar system (GOMT 2000a). 200 sources of dust have been used.
This 3D-model is based on a spherical $45\times 180\times 245$ grid with
$2\cdot 10^6$ cells containing $0.6\cdot 10^{11}$ particle positions.
}
\label{fig2}
\end{figure}
\noindent 
containing $(0.5-0.6)\times 10^{11}$ positions
of dust particles. Here we present the results concerning the distribution
of dust particles (of radius 1-2 $\mu$m) produced by 100 KBOs from both
the pericenter and apocenter of each. Figs.~2a,b show 
the spatial structure of kuiperoidal dust up to 60 AU.
Our simulations reveal a new dust component in the form of the
gravitationally scattered kuiperoidal dust in the belts near Jupiter and
Saturn. This {\it scattered} population is basically non-resonant, it
is highly inclined and possesses large eccentricities. A major part of this
component is ejected from the Solar system while passing
by Saturn's  and  Jupiter's orbits.

The other, {\it resonant} component of kuiperoidal dust is responsible for
the regions of elevated dust density. These resonant dust belts, especially
near the Neptune's orbit, can be seen in Fig.~1b. Our simulations  used
two particle sizes ($1-2~\mu$m and $5-10~\mu$m). The smaller the particle 
size, the smaller is 
\begin{figure}[!ht] 
\centerline{\epsfig{file=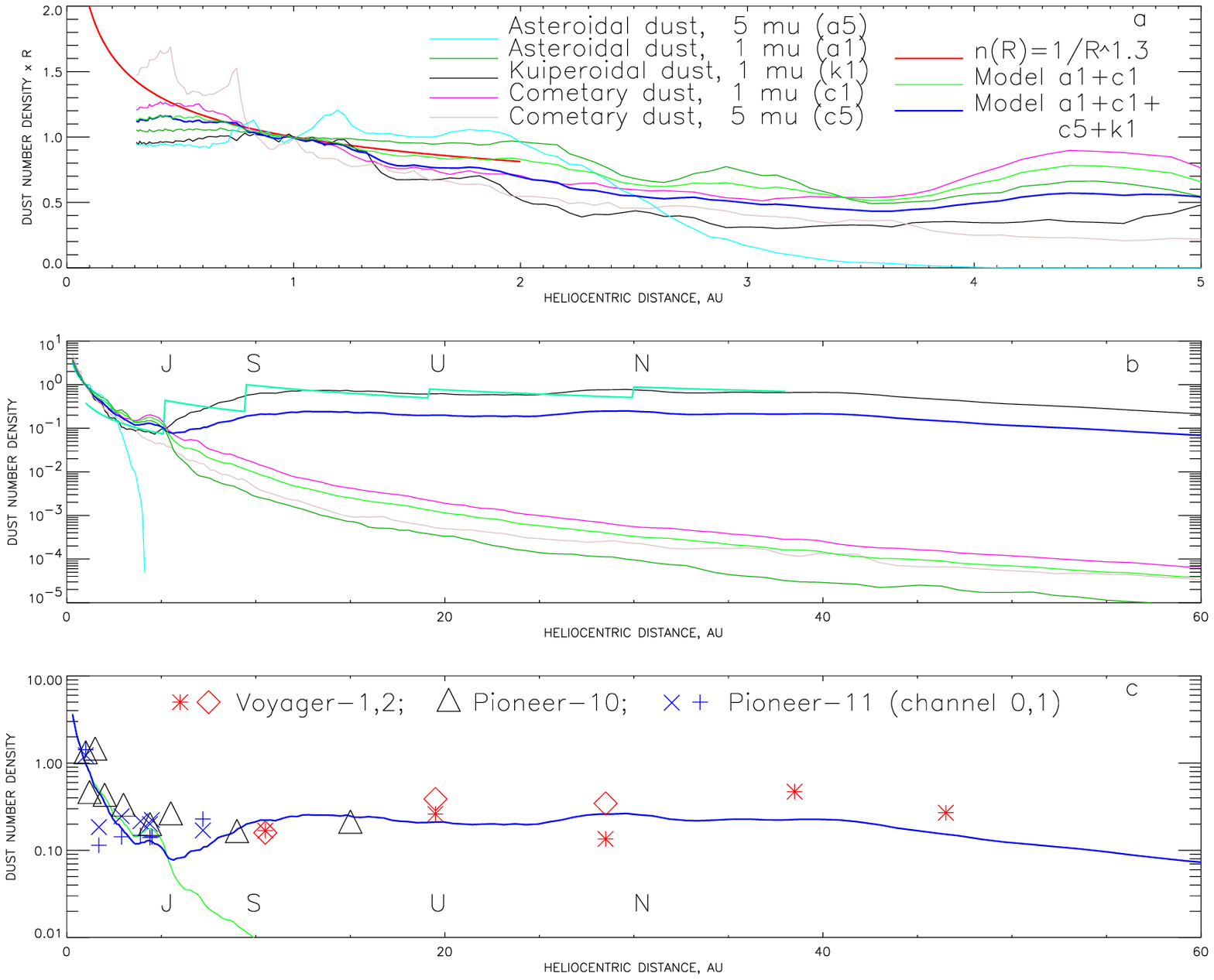,width=5.5in,height=4.7in}}
\caption{\ssq
 The simulated number density of various components of the IPD cloud in the
ecliptic plane as a function of heliocentric distance $R$.\\
{\bf a} -- $0.3-5$ AU. The ordinate is multiplied by $R$
so that the fiducial density run $R^{-1}$ yields a constant. Remarkably,
the number density of large asteroidal grains sharply decreases at $R>2$~AU,
in accordance with our analytical model (GOMT 1997a).\\
{\bf b} --  $0.3-60$ AU distance range. The number density of kuiperoidal
dust remains approximately constant at $10-40$ AU.
The saw line offers a simple explanation to this fundamental result:
the number density  of dust  drifting under the P-R drag toward the
Sun increases as $R^{-1}$, but the planets decrease it by ejecting particles
from the Solar system or putting them to more inclined or eccentric
orbits. For visualization purposes, the influence of the planets
is shown as single jumps, although the actual gravitational scattering is
much more smooth.\\
{\bf c} --  $0.3-60$ AU distance range. Comparison of the simulated dust
density confronted with the dust detectors data indicates that the model
without the kuiperoidal component (grey line) strongly disagrees with the
data, whereas  the model
with the kuiperoidal component (heavy line) is in a fair agreement with them.
}
\label{fig3}
\end{figure}
\noindent
the contrast resonant structure / background, in accordance with Liou \& Zook 
(1999). There is a remarkable density minimum between Mars and Jupiter. This 
minimum, which is seen more clearly in Fig.~2b, is due to the fact that 
Jupiter  either ejects from the Solar system or transfers to more
inclined and eccentric orbits an appreciable part of the dust drifting
toward the Sun. An increase of dust number density in the region between
Mars and Earth is explained by the role of the  P-R drag, which results in
the density run $\propto R^{-1}$. Finally, as a major result, we find that
the number density of kuiperoidal dust increases with heliocentric distance
between 4 and 10 AU but forms a plateau between 10 and 50 AU.              

\subsection*{4.3. Which of Dust Components Prevails in the Solar System?}

Our hypothesis that the kuiperoidal dust dominates in the Solar system
(certainly, in its outer regions) has been verified by confronting it with
available data on the dust distribution in the Solar system, both outer and
inner. Two fundamental facts need to be mentioned: (i) at Earth, the dust
density run $R^{-a}$ has the exponent $a=1.3$ (Divine 1993); and (ii)
between Jupiter and Neptune, the dust density is almost constant (Humes
1980, Gurnett et al. 1997).

Fig.~3 shows, for various possible components of the IPD cloud, the number
density of the simulated dust distribution in the ecliptic plane as a
function of heliocentric distance. Within the distance interval of $0.5-1.5$
AU, an averaged density run $R^{-a}$ has the exponent $a=1.5-1.7$ for the
cometary dust; $a=1.4$ for the kuiperoidal dust, and $a=1.0$ for the
asteroidal component. We find that both the two-component and
three-component models described  above give  us the exponent $a$  very
close to $a=1.3$ observed at Earth. However, the difference between the two
models, only marginal at Earth, becomes very significant at large
heliocentric distances. Fig.~3b demonstrates the simulated dust density run 
for different components of the IPD cloud at heliocentric distances up to 
60~AU. Both the cometary dust and small asteroidal dust decrease their number
densities in the distance range of $10-30$~AU as $R^{-(2.4-3.3)}$, whereas
that of the kuiperoidal dust changes insignificantly (as $R^{-0.05}$). Thus
our hypothesis that the kuiperoidal dust dominates in the outer Solar system
explains the Pioneer and Voyager data
fairly well, while the traditional view that the bulk of the IPD cloud is
produced by the Jupiter family comets turns out in contradiction with the
available data of dust detectors, as is clearly seen in Fig.~3c.

\section*{5. Modelling the Zodiacal Light}

\subsection*{5.1. Fitting the COBE DIRBE Data}

We have computed the contribution to the
zodiacal light from each of the components of the IPD cloud described 
in Secs.~3 and 4.
Figs.~4a,b show the results from these computations. Also shown is
a comparison with some representative DIRBE data on the zodiacal
light, where the comparison is constructed by a weighted mixture of the
various components. As can be
seen from the ratio $I_{ecliptic}/I_{pole}$ shown in Fig.~4a, both the
asteroidal and kuiperoidal components have a flatter shape than the actual
IPD cloud. In contrast, the Jupiter family comets create a cloud thicker
than the actual one.
\begin{figure} [!ht] 
\centerline{\epsfig{file=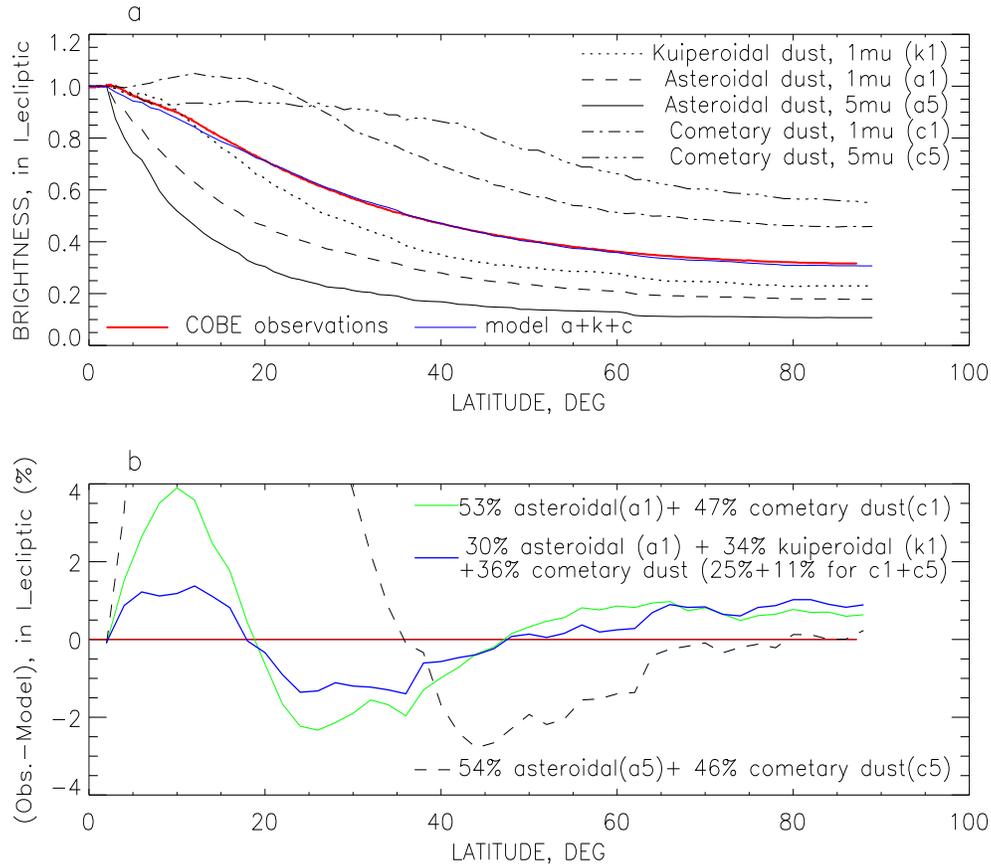,width=5.5in,height=4.7in}}
\caption{\ssq
Infrared emission of different components of the simulated IPD cloud 
({\bf a}) and a detailed comparison of our models with  
the COBE/DIRBE data at $\lambda=25~\mu m$, elongation$=90^\circ$  ({\bf b}).
The asteroidal dust is assumed to be produced by 220 asteroids,
the cometary dust -- by 388 comets, and
the kuiperoidal dust - by 200 sources. The total number of the computed
particle positions is more than $10^{11}$. The
cometary  particles form a thicker cloud, whereas the asteroidal as well as 
kuiperoidal particles -- a thinner cloud than the observed one. 
A maximal / an average deviation from the COBE data is  18\%/7\% for the  
``a5 $+$ c5"  asteroidal-cometary  model and 4\%/1.5\%  for the ``a1 $+$ c1" 
model. All other combinations of (a1, a5, c1, c5) components would only give
intermediate values of the deviations. Meanwhile  the 
``asteroidal-cometary-kuiperoidal" model fits the data much better: 
a disagreement with the data is 0.85\% on average and never exceeds 1.4\%.
}
\label{fig4}
\end{figure}

As seen in Fig.~4b, a three-component (`asteroidal-cometary-kuiperoidal')
physical model of the IPD cloud describes the COBE data with an average
accuracy of 0.85\%. Although this model employs three
free physical parameters (which actually can be determined from observations),
it offers an accuracy 
that is comparable with the best phenomenological model of the
zodiacal cloud using about 50 free parameters (Kelsall et al. 1998).

\subsection*{5.2. Modelling the Zodiacal Light Far from Earth}

Space observations far from Earth, {\it e.g.}, at 3~AU, would offer 
substantial improvements in the zodiacal light emission and scattering 
(Mather and Beichman 1996). Unfortunately, a rather accurate multi-parametric 
model of the zodiacal brightness derived by Kelsall et al. (1998) from the 
COBE data cannot be 
\begin{figure}[!hb] 
\centerline{\epsfig{file=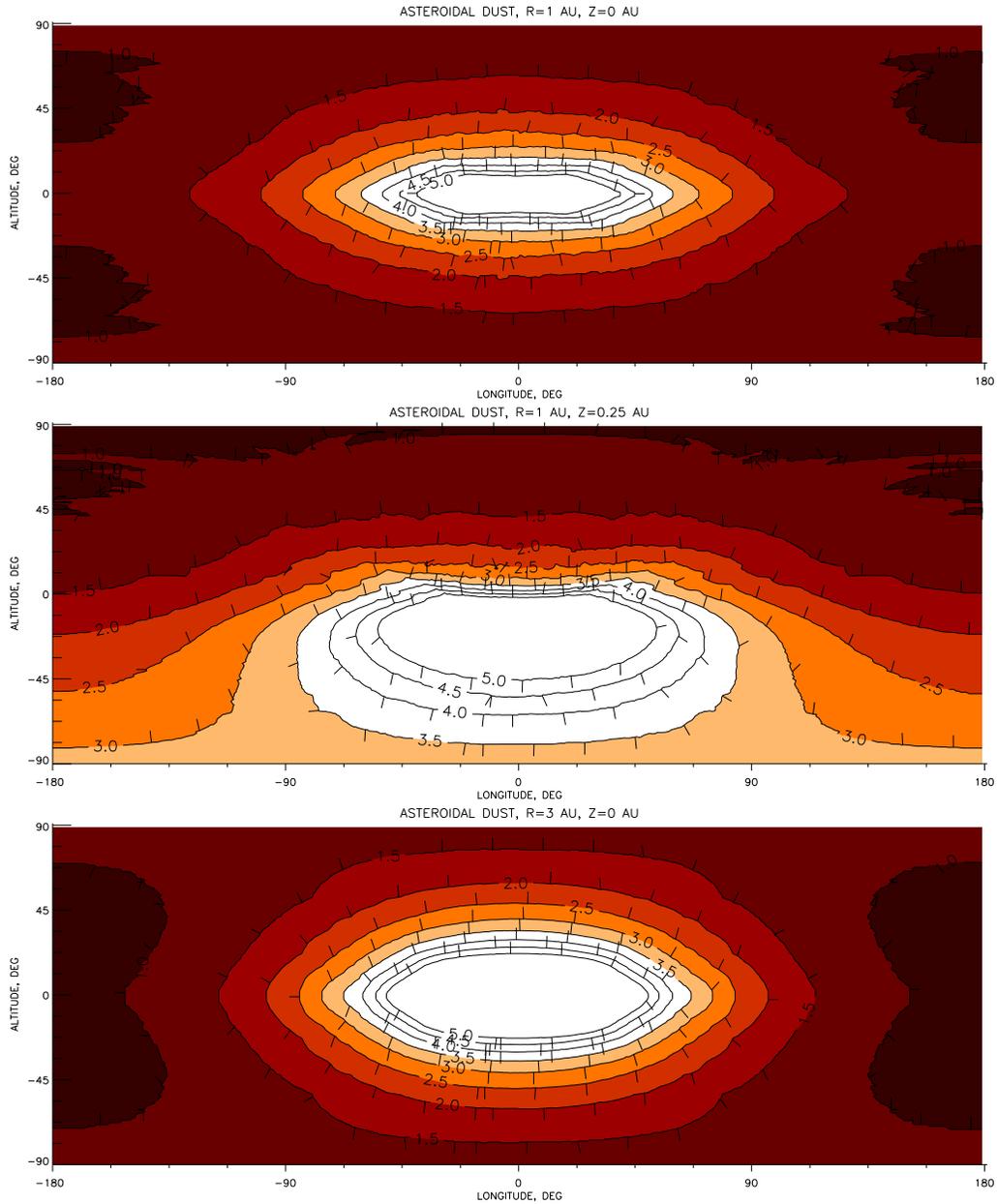,width=5.5in,height=6.5in}}
\caption{\ssq
Emission of a small ($r=1-2\mu$m) asteroidal dust in the 5$\mu$m band
}
\label{fig5}
\end{figure}
\noindent 
reliably extrapolated to heliocentric distances as large 
as 3~AU. Meanwhile our physical modeling of the zodiacal cloud 
makes it possible to evaluate quantitatively the zodiacal light emission 
and scattering throughout the Solar system (GOMT 2000b). 

Using the inferred distribution of the zodiacal dust,
we have computed a variety of zodiacal light maps, both for thermal    
emission and scattered components, at different locations $(R,Z)$ 
of the observer. Fig.~5 gives a representative illustration of the 
brightness of asteroidal dust at 5~$\mu$m as a function of latitude 
$\varphi$ and longitude $\lambda$ (in the
telescope's frame) is given in logarithmic scale (the neighboring contour
intensities differ by ${\sqrt e}$). The Sun's position is
(0,0) for $Z=0$ and is shifted to a negative  $\varphi$  for $Z=0.25$ AU.
At each location, there is a minimum in the zodiacal light which can be seen
as a `dark spot' (or several spots). The positions of 
those minima are explained by an interplay between dependencies of emissivity
upon density and temperature.
  
\subsection*{5.3. Warp of the Zodiacal Cloud}

The smooth zodiacal dust cloud is inclined to the ecliptic plane by $2.03\pm 
0.017^\circ$ (Kelsall et al. 1998). Since during the annual Earth motion the 
bulk of the zodiacal dust is positioned either above the ecliptic plane or 
below it, the zodiacal light turns out to be variable.

The cause of warp of the zodiacal cloud is as yet a puzzle. Interestingly,
a similar warp has been recently discovered in the circumstellar disk around 
Beta Pictoris (Heap et al. 2000 and refs. therein). We point out that the 
observed warp in both cases is caused by the presence of a massive planet 
such as Jupiter. To prove this, we have plotted the North pole emission
using both the data of 41-week helium-cooled period and (virtually so far
unused) the data of a more extended (about 3 years), `warm' period of COBE 
mission (Kelsall et al. 1998).
We find a well-pronounced 12-month period, which indicates that an inclined
part of the zodiacal cloud is stationary in the inertial frame. Therefore,
the warp of the cloud could be associated with the influence
of Jupiter on a non-resonant part of the cloud, and the annual variations
of the zodiacal emission are induced by the Earth orbital motion through
the warp.

Our numerical simulation confirm  the efficiency of this  process. Since
the warp of the disk around $\beta$~Pic can be visualized easier than that 
around the Sun, it would be instructive to address our numerical simulation 
of $\beta$~Pic dust disk to illustrate the warp induced by a Jupiter-like 
planet in an orbit inclined to the disk (see Gorkavyi et al. 2000c). 
As shown there, a comparison of that modeling with the STIS observations 
makes the proposed interpretation quite plausible.
		  
\subsection*{5.4. Simulations of Zodiacal Dust in Circumstellar Disks}

While warping of a dusty disk, like $\beta$~Pic, seen edge-on might serve 
as a signature of an embedded planet, simulations of circumstellar disks seen 
face-on need a different approach. The major factors which cause the
re-distribution of dust visible in the circumstellar disk in the presence of
an embedded planet are: (i) gravitational scattering by the planet, which
produces a central cavity (a `hole') and (ii) resonances, which produce
asymmetry in the dust distribution in the form of clumps, arcs, rings, etc.
As the dust passes by the planets in its infall, it interacts with them by
accumulating in the outer planetary resonances (Liou \& Zook 1999, OGMT 
2000). As we demonstrate in OGMT (2000), the resonant structure in the dusty 
circumstellar disks seen face-on, like Vega or Epsilon Eridani, could serve 
as an efficient tool of planet detection in those systems.

Thus, along with the ability to compute an inproved physical model of the IPD
cloud in the Solar system and to explain some key aspects of the available 
data with it, our tools enable us to predict important details in the dust 
structure near other stars to be tested in future observations. For instance, 
using our modeling, we predict that the resonant
asymmetric feature revolves around $\epsilon$ Eri and Vega with an angular 
velocity measurable within a few years (OGMT 2000, Gorkavyi et al. 2000c).

\section*{7. Future Work}

We plan to improve our physical modeling so as to fit the DIRBE data with a
precision much better than 1\%. To this end, one needs:
\begin{itemize}

\item{} to accurately compute the density distributions of all dust components
   incorporating particles of 5 to 10 different sizes and accounting for
   all known as well as simulated sources of dust; 
 
\item{} to account for evaporation and sputtering of dust as a function of
   heliocentric distance;

\item{} to include the short-term (days to months) variability and small-scale
  phenomena in the zodiacal cloud.                                   
\end{itemize}

The last point deserves a more detailed discussion. The time variations 
found in the zodiacal light are at a few \% level
for the short ($\lax 5~\mu$m) wavelength bands and of order of 0.5 to 1\% for
the longer wavelength bands (Kelsall et al. 1998), {\it i.e.}, appreciably
greater than expected on the basis of known variability in the bolometric 
output from the Sun.                
	    
The revealed short-term variability might contain contributions of
different origins, such as (i) variability of solar wind;
(ii) inhomogeneities in the zodiacal cloud along
  the Earth orbit associated, e.g., with the Earth resonant ring and
  recent meteoroidal dust or cometary tails; and (iii) dust-source dependence
  and/or dust response to UV or particle
  change with time as the solar-wind and/or UV heating 'cook' the particles. 

Given that the IPD inhomogeneities are located at the distances from Earth
not exceeding several AU, their parallaxes are as large as tens of degrees.
This provides a unique opportunity to reconstruct, via a 3-D `computer
tomography', a 3-D map of small-scale inhomogeneities in the zodiacal cloud 
using the 41 week data.

Knowledge of the variability and local structure of the IPD cloud would allow
us to make the next important step and to determine the absolute value of the 
zodiacal emission. Indeed, in a given direction on the sky, the Galactic 
emission is constant, whereas the zodiacal emission varies in time. The 
physical modeling of the emission amplitude in the above direction could 
enable us to disentangle the Galactic and zodiacal contributions.
As a result, a very accurate fitting of  the zodiacal light would be 
possible, able to solve two major problems: (i) to derive the basic physical 
parameters of the zodiacal cloud containing a valuable information 
concerning the structure, dynamics, and origin of this cloud, 
and (ii) the residuals would make it possible to constrain
or even evaluate the contribution of the extragalactic infrared background.

\section*{8. Summary and Conclusions}

We have developed a physical model of the zodiacal cloud incorporating
the real dust sources of asteroidal, cometary, and kuiperoidal origin. 

Using our codes, we have demonstrated their efficiency and power by
performing the following simulations:  

 (i) distribution of the scattered comets, which enables us to reveal
 the four `cometary belts' associated with the orbits of four giant planets, 
 which are expected to contain 20-30 million of cold comets; 
 
 (ii) detailed analysis of a rich resonant structure found in these belts,
 which predicts the existence of gaps similar to the Kirkwood gaps;
 
 (iii) a 3-D physical model of the IPD cloud, which explains the available
 data of Pioneers and Voyagers dust detectors;
 
  (iv) zodiacal light distribution in the Solar system, which fits the COBE
   data with an average accuracy of 0.85\%, and 
  
  (v) resonant structure in dusty circumstellar disks of Vega and Epsilon
  Eridani and a warp in dusty disk of Beta Pictoris considered tobe 
  a signature of embedded extrasolar planets.
      
 Under a set of reasonable assumptions, it seems safe to conclude: 
 
1. The {\it kuiperoidal} dust plays a role more important than previously 
recognized. It appears to account for the space dust observations beyond 
6~AU, while near Earth it could possibly contribute as
much as 1/3 of total number density (1/4 of surface density) and 1/3 of
the zodiacal emission near ecliptic.

2. The two other components of the IPD cloud, the {\it cometary} and 
{\it asteroidal} dust contribute respectively 36\%
and 30\% of the number density  and the zodiacal emission (at ecliptic) near
Earth. The cometary particles contribute 60\% to the surface density  of the
IPD cloud  near Earth. A solely two-component model (i.e. without the
kuiperoidal dust) would give a worse fit of dust distribution at Earth and 
would fail entirely for the outer Solar system.

3. Further simulations of resonances associated with the planets
embedded in  dusty circumstellar disks could enable a breakthrough in the
understanding of the circumstellar disk structure and lead to possible
planet detection long before direct imaging can find them.

Further development of a multicomponent, high-precision (at the level of a 
few $\times$0.1\%) model of the IPD cloud would allow to solve at a new 
qualitative level a number of key astronomical problems:

\begin{itemize}
\item{} to get an important information about the major physical effects
     operating in the Solar system, such as the PR drag, resonant captures,
     gravitation scattering, role of interstellar particles (including their
     collisions with the IPD particles), evaporation of dust, efficiency of
     dust production at different distances from the Sun, etc.;     
						      
 \item{}    to evaluate the parameters of yet undiscovered cometary and
     asteroidal populations contributing to the origin of the zodiacal
     cloud;

\item{} to get a reliable template for exo-zodiacal clouds (circumstellar
     dusty disks) as a basis for revealing the embedded exo-planets;
									 
\item{} to help in interpreting and guiding a number of space missions
    with dust collectors, such as CASSINI and STARDUST; 
									 
\item{} to help in planning the targets for space infrared telescopes,
    such as SIRFT  and NGST;     

\item{} to improve evaluations of micro-meteoroid impacts for spacecraft;      
			     
\item{} to subtract  the zodiacal contribution from the COBE DIRBE data
   with a high precision to evaluate/constrain the Cosmic Infrared 
   Background.
\end{itemize}
\vspace{0.1truein}

\noindent{\bf Acknowledgements}. This work would be impossible without
my collaboration with Nick Gorkavyi, John Mather, and Tanya 
Taidakova. I am thankful to Thomas Kelsall for many fruitful 
discussions. Support by NASA Grant NAG-7065 to George Mason University 
is acknowledged.




\begin{references}
\def\ref{\reference}
\ref{Backman, D.E., Dasgupta, A. \& Stencel, R.E. 1995,
     ApJ 450, L35}
\ref{Dermott, S.F. et al.
1994, Nature 369, 719}
\ref{Dermott, S.F. et al.
1996, in {\it Physics, Chemistry, and Dynamics of Interplanetary Dust},
   ed. $~$ B.~Gustafson \& M. Hanner, 
ASP Conf. Ser. 104, p.~143}
\ref{Divine, N. 1993, J. Geophys. Res. 98E, 17029}
\ref{Flynn, G.J. 1994, Lunar \& Planetary Science XXV, 379}
\ref{Flynn, G.J. 1996, in {\it Physics, Chemistry, and Dynamics
  of Interplanetary Dust}, ed. $~$ B.~Gustafson \& M. Hanner, 
ASP Conf. Ser. 104, p. 171}
\ref{Gorkavyi, N.N., Ozernoy, L.M. \& Mather, J.C. ($\equiv$ GOM) 1996, in
   {\it Physics, Chemistry, and Dynamics of Interplanetary Dust},
   ed. B. Gustafson \& M. Hanner, (San Francisco: ASP),
   ASP Conf. Ser. 104, p. 43}
\ref{Gorkavyi, N.N., Ozernoy, L.M. \& Mather, J.C. ($\equiv$ GOM) 
1997, ApJ 474, 496}
\ref{Gorkavyi, N.N., Ozernoy, L.M., Mather, J.C. \& Taidakova, T.
($\equiv$ GOMT) 1997a, ApJ 488, 268}
\ref{Gorkavyi, N.N., Ozernoy, L.M., Mather, J.C. \& Taidakova, T.
1997b, Bull. AAS 29, 782}
\ref{Gorkavyi, N.N., Ozernoy, L.M., Mather, J.C. \& Taidakova, T.
1997c, Bull. AAS 29, 1310}
\ref{Gorkavyi, N.N. et al. ($\equiv$ GOMT) 1998a, Earth, Planets and Space, 
50, 539}
\ref{Gorkavyi, N.N., Ozernoy, L.M., Mather, J.C. \& Taidakova, T.,
 1998b, BAAS 30, 853}
\ref{Gorkavyi, N.N., Ozernoy, L.M., Mather, J.C. \& Taidakova, T.,
1998c, BAAS 30, 1143}
\ref{Gorkavyi, N.N. et al.  ($\equiv$ GOMT) 
2000a, {\tt astro-ph/0006435}; Planetary Space. Sci. (submitted)}
\ref{Gorkavyi, N. et al. ($\equiv$ GOMT) 
2000b,
{\it The NGST Science and Technology Exposition} (E.P.~Smith and K.S.~Long,
eds.) ASP Conf. Ser. 207, 462}
\ref{Gorkavyi, N.,  Ozernoy,  L., Mather, J., \& Heap, S.
2000c, In {\it Disks, Planetesimals,
and Planets} (F.~Garzon et al., eds.)
ASP Conf. Ser. (in press); WWW e-print {\tt astro-ph/0005347}}
\ref{Gurnett, D.A. et al. 1997, Geophys. Res. Lett. 24, p.~3125}
\ref{Haug, U. 1958, Zeitschrift f\"ur Astrophysik, 44, 71}
\ref{Hauser, M.G. et al.
1998, ApJ, 508, 25}
\ref{Heap, S. et al. 2000, Ap.J. (in press)}
\ref{Humes, D.H. 1980, J. Geophys. Res., 85, 5841}
\ref{Jackson, A.A. \& Zook, H.A. 1989, Nature 337, 629}
\ref{Jewitt, D. 1999, Ann. Rev. Earth. Planet. Sci., 27, 287
}
\ref{Kelsall, T. et al.
1998, ApJ, 508, 44}
\ref{Kessler, D.J. 1981, Icarus 48, 39}
\ref{Leinert, L., Roser, S., \& Buitrago, J. 1983, A\&A, 118, 345}
\ref{Levison, H.F. \& Duncan M.J. 1997, Icarus 127, 13}
\ref{Liou, J.-C. \& Zook, H.A. 1999, Astron. J. 118, 580}
\ref{Liou, J.-C., Zook, H.A. \& Dermott, S.F. 1996, in
{\it Physics, Chemistry, and Dynamics of Interplanetary Dust},
    ed. B. Gustafson \& M. Hanner, (San Francisco: ASP),
ASP Conf. Ser. 104, p. 163; Icarus 124, 429}
\ref{Malhotra, R., Duncan, M., \& Levison, H. 1999. In {\it Protostars
 and Planets IV} (in press) $=$ astro-ph/9901155}
\ref{Mather, J.C. \& Beichman, C.A. 1996, {\it Unveiling the Cosmic Infrared
Background}, Ed. E.~Dwek, AIP Conf. Proc. 348, 271}
\ref{Ozernoy, L.M., Gorkavyi, N.N., \& Taidakova, T. 2000a,  
Planetary Space Science, 48, 993}
\ref{Ozernoy, L.M., Gorkavyi, N.N., \& Taidakova, T.  ($\equiv$ OGT)  2000b,
   Mon. Not. R.A.S. 2000 (submitted), an early version posted in 
    {\tt  astro-ph/9812479}}
\ref{Ozernoy, L.M. et al. ($\equiv$ OGMT),
2000c, Astrophys. J. 537, L147}
\ref{Stern, A. 2000, in {\it Highlights of Astronomy, JD 4}, A.~Lemaitre
\& H.~Rickman, eds. ASP Conf. Ser. (in press)}
\ref{Taidakova, T. 1997, in {\it Astronomical Data Analyses,
 Software and Systems VI}, ed. G. Hunt \& H.E. Payne, 
ASP Conf. Ser. 125, p. 174}
\ref{Taidakova, T. \& Gorkavyi, N.N. 1999, {\it
    The Dynamics of Small Bodies in the Solar Systems: A Major Key to
    Solar Systems Studies}, Eds.  B.A. Steves and B.A. Roy,  Kluwer
    Academic Publishers, 393
}
\ref{Valsecchi, G.B. \& Manara, A. 1997, A\&A, 323, 986}

\end{references}
\end{document}